\documentclass[10pt,conference]{IEEEtran}
\IEEEoverridecommandlockouts
\usepackage{amsmath,amssymb}
\usepackage{booktabs}
\usepackage{graphicx}
\usepackage{microtype}
\usepackage{cite}
\usepackage{url}
\usepackage[hidelinks]{hyperref}

\title{Precision-Preserving Subtensor Inference for a 22B Joint Audio--Video Generator under a Fixed GPU Working Set}
\author{\IEEEauthorblockN{Hanyue Shen and Xinling Liao}
\IEEEauthorblockA{YH Intelligence Technology, Co., Ltd.\\
Beijing, China\\
hanyueshen@yh-intel.com; xinlingliao@yh-intel.com\\
ORCID: H. Shen 0009-0009-5934-6384; X. Liao 0009-0000-3137-2825}
\thanks{Accepted to the CONSTRAINT 2026 Workshop, held in conjunction with IEEE BigData 2026.}}

\begin{document}
\maketitle

\begin{abstract}
Complete-block offloading reduces model residency but requires each transferred block to fit the GPU budget. We present Precision-Preserving Subtensor Inference (PSI), a fixed-arena runtime that combines sub-block weight streaming with separately managed intermediate state for the BF16 LTX-2.5 22B joint audio--video transformer. The contribution is a construction whose managed GPU footprint is determined by reusable pages, live state, and bounded workspace, together with controls that separate resource use, transport correctness, and numerical fidelity. Its partitioned operators preserve the dense real-arithmetic construction under stated assumptions. Across nine matched full-prompt cases on RTX A6000, modified PSI reduces sampled transformer-process peak memory from 8,578 to 360 MiB relative to official MATH, while mean time for the eleven transformer invocations rises from 247.11 to 349.00 s. Its managed arena is 80 MiB. Complete native generation and a complete-block FLASH path both succeed under a fixed 3 GiB available-memory cap. The matched A6000 comparison and separate DGX Spark service suite both achieve 100\% success in complete audiovisual decoding. FFN reuse lowers mean block time by 4.0\% with unchanged BF16 outputs; a fixed-arithmetic one-token transport control preserves all saved states bitwise. Targeted bias and attention changes improve local and one-token numerical comparisons, while mean endpoint error increases across nine full-prompt cases. The construction enables BF16 audiovisual inference with an explicit transformer-memory budget and makes numerical fidelity a separate, testable property of the implementation.

\end{abstract}

\begin{IEEEkeywords}
resource-constrained inference, diffusion transformer, GPU memory, weight streaming, audiovisual generation
\end{IEEEkeywords}

\section{Introduction}

Large generative models face two GPU-memory constraints: the checkpoint may exceed accelerator capacity, and the weights and state needed by one operation or block may exceed the budget assigned to a request. Complete-block offloading addresses model residency while retaining the block as the transfer and execution unit. PSI schedules operations inside that block using smaller weight slices and stores inactive intermediate state on the host. Joint audio--video generation makes this distinction useful because conditioning, transformer execution, and media decoding have different live-state requirements. Lightricks' LTX-2.5 setup guide lists at least 32 GB of GPU memory and recommends an A100 (80 GB) or H100 \cite{ltx25requirements}.

Prior systems manage activations, tensor placement, and tile reuse across the memory hierarchy \cite{vdnn,swapadvisor,zeroinfinity,flexgen,welder}. PSI applies these established techniques to a BF16 multimodal diffusion transformer through a fixed device arena and an indexed host journal. The execution question is how to schedule each operator within that arena. The measurement question is how this managed working set relates to the transformer process and the complete generation pipeline.

Our prototype targets the official distilled BF16 LTX-2.5 22B checkpoint \cite{ltx25model}, a successor to the joint audio--video LTX-2 model \cite{ltx2}. It declares the BF16 write boundaries and execution dependencies used by each operator. Matched controls separate transport scheduling from numerical evaluation: a transport-only comparison fixes the arithmetic, while baseline and modified PSI are compared with the official MATH transformer on one-token and full-prompt inputs. The modified variant evaluates targeted numerical changes to bias addition and attention accumulation.

The paper makes four contributions:
\begin{itemize}
  \item We construct a fixed-arena runtime that combines reusable weight pages with indexed host state. Real-arithmetic identities establish the partitioned operators, and explicit event dependencies support byte-preserving transport.
  \item We measure an 80 MiB managed arena and a 360 MiB transformer-process peak across eighteen full-prompt native trajectories. Allocation traces, capacity probes, and complete-block controls explain these resource scopes.
  \item We separate numerical and scheduling effects through same-input operator probes, matched trajectories, and repeated block measurements. These controls identify floating-point differences and test optimizations with arithmetic held fixed.
  \item We demonstrate complete generation under a fixed 3 GiB available-memory cap, with a successful FLASH control. The matched A6000 study produces 27/27 successfully decoded outputs, and a separate Spark GB10 service suite produces 30/30; media similarity and prompt adherence are evaluated separately.
\end{itemize}

\section{Design}

\subsection{Managed working-set contract}

Let $S_{\mathrm{live}}$ be the reserved capacity for the current operator's state and scratch, including alignment and unused capacity, $W$ a fixed reserved library workspace, $P$ the maximum page size, and $\epsilon$ bounded metadata. With two reusable pages, the declared managed working set is
\begin{equation}
M_{\mathrm{managed}} \leq S_{\mathrm{live}} + W + 2P + \epsilon.
\label{eq:bound}
\end{equation}
Equation~\eqref{eq:bound} accounts for PSI-managed device allocations for a selected workload. Independence from checkpoint size requires parameter consumers partitionable into pages of at most $P$, bounded state/scratch reservation and library workspace, and page reuse ordered after the last consumer. Device-side journal staging buffers are included in $S_{\mathrm{live}}$ or $W$. Host storage, total host RSS, CUDA context, external allocations, and other audiovisual pipeline phases lie outside this contract; its conditions must be checked for each operator and token length.

\begin{figure*}[t]
\centering
\includegraphics[width=.96\textwidth]{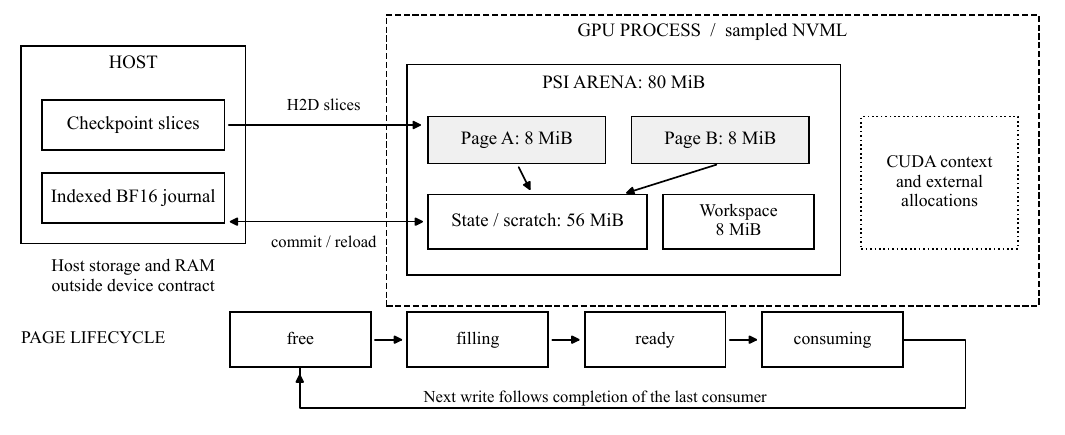}
\caption{Fixed-arena construction. Checkpoint slices enter reusable pages; state is committed to and reloaded from indexed host buffers. The next write to a weight page follows completion of its last consumer. The arena is inside the GPU process; external allocations and CUDA context are outside its 80 MiB accounting. Box areas are schematic.}
\label{fig:construction}
\end{figure*}

Four rules define the contract: PSI-managed device allocations come from the arena; each weight page has at most one unreleased slice request; reuse follows the required copy and consumer dependencies; and operators declare their BF16 write boundaries. In the measured configuration, two 8 MiB pages and state/workspace/scratch share an 80 MiB arena. Figure~\ref{fig:construction} shows the allocation boundary and page lifecycle.

\subsection{Exact arithmetic equivalence}

For fixed checkpoint values and scalar functions, subtensor evaluation preserves the dense construction over the reals. Let $X\in\mathbb{R}^{N\times d}$, $U\in\mathbb{R}^{h\times d}$, $V\in\mathbb{R}^{r\times h}$, with biases $a,b$ and a complete, disjoint hidden-channel partition $\{J\}$. For the coordinatewise tanh-GELU function $\phi$,
\begin{equation}
 \begin{aligned}
 H_J&=\phi(XU_{J,:}^{\mathsf T}+\mathbf{1}a_J^{\mathsf T}),\\
 F(X)&=\sum_J H_JV_{:,J}^{\mathsf T}+\mathbf{1}b^{\mathsf T}.
 \end{aligned}
 \label{eq:ffn}
\end{equation}
Each hidden coordinate receives its complete dot product and bias before $\phi$. Concatenating the $H_J$ therefore recovers the dense hidden tensor; the disjoint down-projection sum recovers every output entry, with output bias added once. Computing each $H_J$ once and reusing it preserves these operands.

For one attention query with finite active scores $s_j=\alpha q^{\mathsf T}k_j$, online softmax retains the following invariant after a nonempty processed key set $A$:
\begin{equation}
 \begin{aligned}
 m&=\max_{j\in A}s_j,\qquad \ell=\sum_{j\in A}e^{s_j-m},\\
 z&=\sum_{j\in A}e^{s_j-m}v_j.
 \end{aligned}
\end{equation}
For the next disjoint tile $T$, set $m'=\max(m,\max_{j\in T}s_j)$ and $\eta=e^{m-m'}$, then update
\begin{equation}
 \begin{aligned}
 \ell'&=\eta\ell+\sum_{j\in T}e^{s_j-m'},\\
 z'&=\eta z+\sum_{j\in T}e^{s_j-m'}v_j.
 \end{aligned}
\end{equation}
The identity $e^{m-m'}e^{s_j-m}=e^{s_j-m'}$ proves the invariant by induction from the first nonempty tile. Since $\ell>0$, $z/\ell$ equals dense softmax attention for any complete key partition. This established IO-aware construction \cite{flashattention} applies here to unmasked, noncausal, dropout-free attention with at least one key per active row.

Full-row RMS replay computes $y_i=\gamma_i x_i/\sqrt{s/d+\varepsilon}$ using the global $s=\sum_J\sum_{i\in J}x_i^2$, total width $d$, and one positive denominator for every tile. Split RoPE transforms $(a,b)=(x_j,x_{j+d/2})$ to $(ca-sb,sa+cb)$ when the original pair values, table indices, and layout are retained. These finite-sum and pair identities establish the real-arithmetic construction.

\subsection{Floating-point differences}

Floating-point implementations evaluate these identities through finite-precision reductions and intermediate writes. Changing a reduction order or a BF16 write boundary can change the result. For BF16 round-to-nearest, ties-to-even $B$, let $z=257/256$ and $b=1/256$. Then
\begin{equation}
 B(B(z)+b)=1\ne129/128=B(z+b).
\end{equation}
The extra BF16 write changes the result although the real-arithmetic expressions are identical. This mechanism motivates the implementation choices below. Actual projection operands and matched trajectories quantify their numerical effects in Section~\ref{sec:numerical}.

\subsection{Ownership and journaling}

A weight page is assigned to one outstanding slice request at a time. Its conceptual lifecycle is \emph{free} (available for writing), \emph{filling} (host-to-device copy in progress), \emph{ready} (copy complete), and \emph{consuming} (used by computation). Copy completion permits consumption; completion of the last consumer permits the next write. Stream ordering and cross-stream event waits enforce these dependencies. The full-prompt implementation copies weights on the compute stream. In the asynchronous weight streamer, release records an event after all uses have been enqueued and ends the request's ownership; the next copy waits for that event to complete before overwriting the page. Recording the event therefore schedules a dependency while device work may still be in progress.

Intermediate state follows a separate path. An indexed host journal stores BF16 values by block, branch, and tensor role. The device-to-host copy waits for its producer; host commit waits for that copy to finish before moving the staged bytes into their journal location. Device-buffer overwrites and later host-to-device reloads retain the previous-consumer dependencies. A later operator reloads the same stored BF16 values.

For a fixed deterministic arithmetic dependency graph, byte-preserving transport with these dependencies preserves outputs: in topological order, every operation receives the same operand bytes and computes the same result. Producer completion precedes consumption, and the last consumer completes before its buffer is overwritten.

\section{Implementation and method}

\subsection{Numerical implementation choices}

Both native variants write the biased up projection as BF16 before GELU, then write GELU's output as BF16. Baseline PSI also rounds the audio matrix-product output to BF16 before adding bias; modified PSI adds that bias in FP32 before the pre-GELU BF16 write. Modified PSI also replaces BF16 numerator exponential weights with FP32 online accumulation for 64-dimensional attention heads, changing the reduction schedule. These two changes are evaluated together in trajectories and separately in same-input operator probes.

\subsection{Data flow and scheduling}

A weight slice enters a device page, the operator consumes it, and intermediate state is retained on device or stored in the host journal for later use. Three scheduling optimizations act on this data flow. FFN reuse computes each 1,024-channel hidden tile once, journals its BF16 values, and reuses them across down-projection output tiles. This removes repeated hidden computation while retaining the ordered down-projection operations.

Asynchronous journal transport stages state through a pinned host ring, allowing copies to overlap independent computation while event waits preserve producer and consumer order. Keeping updated state resident removes redundant transfers. Weight prefetch prepares upcoming slices in two alternating pages; computation waits for the required copy, and the next page write waits for its previous consumer. The journal and prefetch variants are tested separately on top of FFN reuse.

\subsection{Workloads and comparisons}

The native CUDA prototype targets LTX-2.5 22B distilled with BF16 weights throughout. Each transformer invocation in the full-prompt cohort contains 48 blocks and separate 1,024-row video and audio conditioning contexts. Stage 1 uses 1,232 video-state tokens; Stage 2 uses 4,928. Both use 126 audio-state tokens. The paired block studies use a first-block input with 4,928 video-state tokens, 126 audio-state tokens, and a frozen one-token text context. Their performance measurements apply to that block input.

Table~\ref{tab:scope} summarizes the experiments, and Appendix~\ref{app:provenance} provides supporting measurements.

\begin{table*}[t]
\caption{Experiment overview. All studies use RTX A6000 except the Spark service suite.}
\label{tab:scope}
\centering\small
\setlength{\tabcolsep}{4pt}
\begin{tabular}{@{}p{0.22\textwidth}p{0.43\textwidth}p{0.31\textwidth}@{}}
\toprule
Experiment & Setup & Measurement \\
\midrule
Transformer memory & 18 native full-prompt trajectories; both stages & Managed arena and process peak \\
One-token replay & Official MATH, baseline PSI, modified PSI; 8+3 steps & Trajectory error, time, and memory \\
Full-prompt comparison & 3 prompts $\times$ 3 seeds $\times$ 3 transformer paths & Numerical fidelity and media validation \\
Full-pipeline budget & Native PSI and FLASH; 3 GiB physical free pool & Completion, time, and memory \\
Transport control & One-token replay; fixed arithmetic & Bitwise equality of saved states \\
Scheduling & 4 block variants; 5 measured runs each & Block time and bitwise equality \\
Capacity and block offload & 6 arena/page settings; official block controls & Feasibility, memory, and numerical error \\
Spark service & DGX Spark GB10; 3 runs $\times$ 10 requests; single slot & Media validation, throughput, and adherence \\
\bottomrule
\end{tabular}
\end{table*}

Each local candidate compares eight stage tensors and the block's final video and audio outputs, giving ten BF16 probes.

Matched trajectories share conditioning, initial state, random tensors, sigma schedules, and preprocessing/upsampling functions; subsequent states evolve separately in each implementation. Same-input operator comparisons are analyzed separately. In the one-token replay, all eleven official denoising states reproduce the frozen BF16 reference values.

The full-prompt cohort fixes three service prompts---a spokesperson, a chef and an airliner---and three unsigned 64-bit request seeds per prompt before execution. The seeded repetitions are labeled run \#1, run \#2, and run \#3 within each prompt category. All 27 trajectories use 1,024-row native projected contexts shared with the official MATH transformer; valid tokenizer lengths are 87, 76 and 77, respectively, with the remaining rows padded. Native text conditioning and the common native media decoders are held fixed to isolate the transformer comparison.

The full-pipeline budget control uses spokesperson run \#1 and the same request seed. We compare official FLASH followed by native PSI under a fixed 3 GiB physical free pool. A direct CUDA allocation reserves the rest of the GPU before each run, leaving an initial physical free pool within 2 MiB below this budget. The reservation stays fixed through tokenizer and native conditioning, regenerated initial/noise tensors, the 8+3 transformer schedule, shared preparation and upsampling, native decoding, and media validation. 

All prompt-dependent stages run again under the reservation; input bytes are checked against the matching full-prompt case. Each path competes for this pool with its CUDA context and library allocations; the reservation helper's own context is reported separately. Separate allocation probes verify small allocations and over-budget rejection. FLASH uses complete-block CPU staging and checked zero-mask elision without a MATH fallback. This is an A6000 available-memory emulation with one attempt per setting.

\section{Results}

\subsection{Memory scopes}

Across the nine full-prompt cases, baseline and Modified PSI each complete an 8+3 trajectory, giving eighteen native trajectories. Every transformer invocation uses an 80 MiB arena and reaches a sampled per-process NVML peak of 360 MiB. These measurements cover both denoising stages, including the larger Stage-2 video state, and describe the transformer process; complete-pipeline memory is measured separately below.

Table~\ref{tab:fullprompt_resources} pairs time and memory for the same nine full-prompt cases. Modified PSI uses 95.8\% less sampled transformer-process memory than official MATH, with 41.2\% greater mean summed invocation time. The common outer timer and equal measurement pauses support this descriptive comparison; cases use a fixed implementation order.

\begin{table}[t]
\caption{Matched full-prompt RTX A6000 time--memory comparison, nine completed trajectories per path. Wall time sums eleven invocations, including startup, input/output, and a 1.5 s measurement pause per invocation. Conditioning, state updates, upsampling, and decoding are outside this timer. Sample SD describes variation across cases. Each path's sampled process peak is identical across its nine cases.}
\label{tab:fullprompt_resources}
\centering\footnotesize\setlength{\tabcolsep}{3pt}
\begin{tabular}{lrr}
\toprule
Path & Mean $\pm$ SD (s) & Peak (MiB) \\
\midrule
Official MATH & $247.11\pm11.65$ & 8578 \\
Baseline PSI & $349.77\pm9.16$ & 360 \\
Modified PSI & $349.00\pm3.22$ & 360 \\
\bottomrule
\end{tabular}
\end{table}

A separate first-block CUPTI trace observes four simultaneously live dynamic device allocations: 80 MiB, 8 MiB, 128 KiB, and 1 KiB, totaling 88.126 MiB. The three smaller allocations precede arena creation; source order associates them with library initialization, while the trace itself records allocation sizes and lifetimes. This dynamic-allocation total remains distinct from the sampled process NVML peak.

\begin{figure}[t]
\centering
\includegraphics[width=\columnwidth]{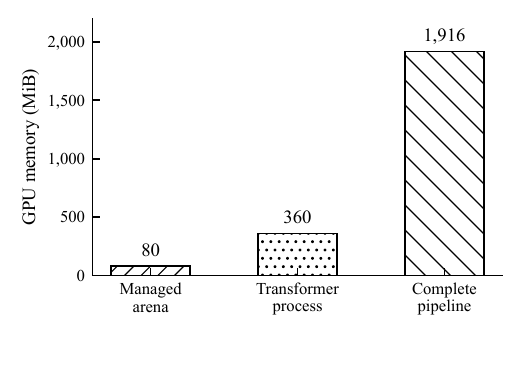}
\caption{RTX A6000 memory scopes. The 80 MiB managed arena and 360 MiB transformer-process peak apply to both denoising stages of all eighteen full-prompt native trajectories. The 1,916 MiB sampled workload-process peak describes complete native generation under the fixed 3 GiB available-memory cap; the FLASH path has the same measured peak. The reservation helper is excluded.}
\label{fig:memory}
\end{figure}

The reserved 80 MiB arena consists of two 8 MiB pages, 8 MiB declared workspace, and 56 MiB scratch; the full-prompt native trajectories reach a maximum of 52.94 MiB scratch use. Figure~\ref{fig:memory} compares their managed arena and transformer-process peak with the full-pipeline process peak under the fixed 3 GiB budget.

\subsection{Numerical discrepancies and targeted changes}
\label{sec:numerical}

\begin{figure*}[t]
\centering
\includegraphics[width=.96\textwidth]{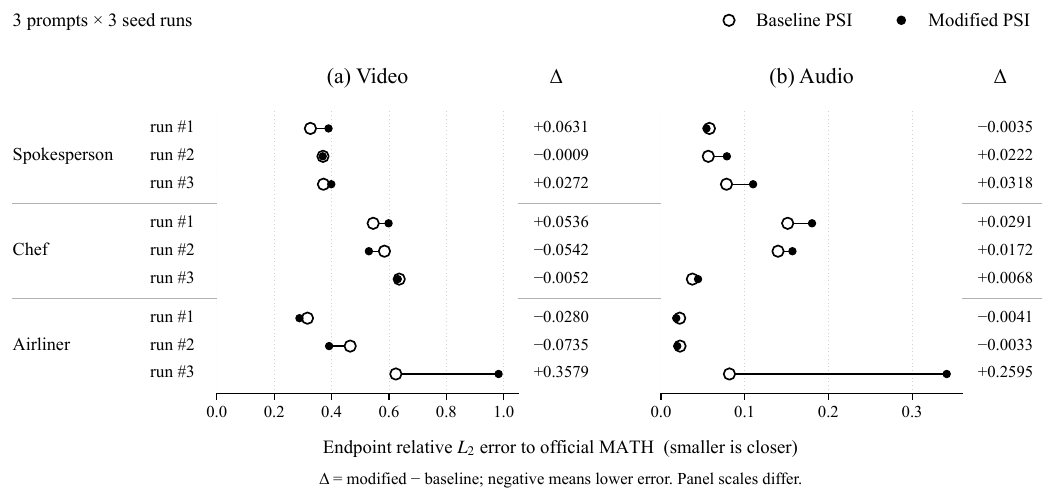}
\caption{Endpoint relative $L_2$ errors against the official MATH transformer for three prompts and three seed runs per prompt on RTX A6000. Each row pairs baseline PSI (open circle) with modified PSI (filled circle); smaller values indicate closer agreement. $\Delta$ is modified minus baseline error. Video and audio use different horizontal scales. All paths share native projected conditioning, initial state, and random inputs within each case.}
\label{fig:full_prompt_numerics}
\end{figure*}

All nine full-prompt cases complete the official MATH, baseline PSI, and modified PSI trajectories. Figure~\ref{fig:full_prompt_numerics} shows every endpoint comparison. Modified PSI lowers relative $L_2$ error in 5/9 video cases and 3/9 audio cases; the mean paired changes (modified minus baseline PSI) are $+0.0378$ and $+0.0395$. The changes improve individual cases while increasing the mean endpoint error in both modalities. All compared endpoint tensors are finite. Independent CPU decoding of the BF16 values reproduces the trajectory error metrics.

The fixed one-token diagnostic (random seed 185) gives a different outcome. All eleven official states reproduce the frozen reference bitwise. The native variants retain an 80 MiB explicit application arena and a 360 MiB sampled process NVML peak. Baseline PSI ends at video/audio relative $L_2$ errors of 0.5057/0.3339; Modified PSI ends at 0.4262/0.2739 (Table~\ref{tab:trajectory_new}).

The same-input diagnostics identify concrete numerical boundaries. Fusing FP32 bias addition before the audio FFN up-projection's BF16 write eliminates 267,593 differing elements among 1,032,192 output elements in the full-shape projection probe. Switching the 64-dimensional-head attention implementation from BF16 numerator exponential weights to FP32 online accumulation reduces the tested audio-self-attention relative $L_2$ error from $1.28\times10^{-3}$ to $1.26\times10^{-5}$, and differing BF16 elements from 68,979 to 45 among 258,048 output elements. The fallback also changes the reduction schedule. These are operator comparisons on identical inputs; the full trajectory includes both changes and subsequent state evolution.

Actual projection operands also expose a finite-precision evaluation discrepancy. At the first mismatching Q coordinate in the baseline PSI diagnostic at denoising state 5 and the first transformer block, with 256-row tiles and 512 output columns per group, the initial normalized/affine tensor matches the official input bitwise. Exact rational evaluation of all 4,096 BF16 products and the bias gives
\begin{equation}
 \begin{aligned}
 y_*&=-\frac{46489387}{2^{39}},\\
 y_{\mathrm{native}}&=-\frac{177}{2^{21}},\qquad
 y_{\mathrm{official}}=-\frac{175}{2^{21}}.
 \end{aligned}
\end{equation}
The native value is the nearest BF16 value to $y_*$; the official projection implementation produces a different BF16 value. Independent integer decoding of the original operands reproduces this result. Thus, identical stored inputs and a common real-valued formula can produce different actual outputs through finite-precision evaluation. Matching a particular official implementation is a distinct target from rounding the exact real sum once.

\begin{table}[t]
\caption{Matched one-token 8+3 replay on RTX A6000. Peak memory is sampled per-process NVML. Invocation wall sums all eleven transformer subprocesses, including initialization, output writing, and a measurement pause of 0.5 s/call for official and 1.5 s/call for native.}
\label{tab:trajectory_new}
\centering\footnotesize\setlength{\tabcolsep}{3pt}
\begin{tabular}{lrrr}
\toprule
Metric & Official & Baseline PSI & Modified PSI \\
\midrule
Invocation wall (s) & 227.385 & 306.352 & 302.163 \\
Process peak (MiB) & 8578 & 360 & 360 \\
Final video rel. $L_2$ & 0 & 0.5057 & 0.4262 \\
Final audio rel. $L_2$ & 0 & 0.3339 & 0.2739 \\
\bottomrule
\end{tabular}
\end{table}

Internal timer intervals differ: the official timer spans input loading through output serialization, while the native timer spans the 48 blocks and final projection.

\subsection{Repeated scheduling measurements}

We repeat the fixed first-block scheduling studies on a GPU reserved for the benchmark. The input has 4,928 video tokens, 126 audio tokens, and one-token conditioning. Four scheduling variants with identical arithmetic run once for warm-up and then for five measured cycles, rotating execution order each cycle. The ten BF16 output tensors match the scheduling baseline in all 24 runs, including warm-ups. Table~\ref{tab:repeats} reports native block-wall means and sample standard deviations; these timers include eight intermediate-output writes and exclude process initialization and the post-work measurement pause.

\begin{table}[t]
\caption{Repeated RTX A6000 first-block scheduling study. Each row has five measured samples; every run preserves all ten test tensors bitwise. Asynchronous journal transport and weight prefetch separately extend FFN reuse.}
\label{tab:repeats}
\centering\footnotesize
\begin{tabular}{lrr}
\toprule
Variant & Mean (s) & Sample SD (s) \\
\midrule
Scheduling baseline & 1.5428 & 0.0188 \\
FFN reuse & 1.4816 & 0.0296 \\
Reuse + async. journal & 1.4349 & 0.0156 \\
Reuse + weight prefetch & 1.4753 & 0.0357 \\
\bottomrule
\end{tabular}
\end{table}

FFN reuse lowers the mean block time by 4.0\% relative to the scheduling baseline. Adding asynchronous journal transport lowers the mean by a further 3.2\%; adding weight prefetch gives a 0.43\% reduction, which is small relative to the observed dispersion. The asynchronous-journal variant keeps updated state resident and omits redundant transfers, reducing activation H2D traffic and host-journal writes by 163,545,088 bytes each per block in the repeated measurements. Its blocking synchronizations decrease from 422 to 297.

To isolate transport over a full trajectory on the fixed one-token input, we hold baseline PSI arithmetic fixed and replace asynchronous journal transport with a synchronous compute-stream implementation. All twelve saved video/audio states across the 8+3 run, including the inter-stage state, are bitwise equal between the two versions.

\subsection{Capacity sweep and complete-block control}

The Modified PSI first-block construction is tested at six discrete arena/page capacities with head, row, output, and FFN tiles fixed. Three configurations complete and produce the same ten tensors bitwise (Table~\ref{tab:capacity}). The smaller arenas exhaust scratch; reducing each page to 4 MiB fails when an FFN weight tile no longer fits. Increasing capacity without changing the computation leaves the successful block times near 1.45 s. These are single capacity probes of this fixed construction.

\begin{table}[t]
\caption{RTX A6000 capacity probes on the 4,928/126-token block test input. Page capacity is per page; there are two pages. Successful outputs match the 80/8 MiB configuration bitwise.}
\label{tab:capacity}
\centering\footnotesize\setlength{\tabcolsep}{4pt}
\begin{tabular}{rrrl}
\toprule
Arena & Page & Wall (s) & Outcome \\
(MiB) & (MiB) & & \\
\midrule
48 & 8 & -- & Scratch exhausted \\
64 & 8 & -- & Scratch exhausted \\
80 & 8 & 1.4458 & Complete \\
96 & 8 & 1.4481 & Complete \\
80 & 4 & -- & Weight tile exceeds page \\
96 & 16 & 1.4497 & Complete \\
\bottomrule
\end{tabular}
\end{table}

The complete-block control stages the same original checkpoint parameters on CPU, moves the unmodified official block and inputs to CUDA, and uses the original SDPA math backend. It runs out of memory at PyTorch allocator caps of 80, 89, 1,024, and 4,096 MiB, and completes at 8,192 MiB with outputs bitwise equal to the official block reference. The 4,096 MiB failure occurs in dense SDPA's 2.90 GiB allocation; the 80 and 89 MiB failures occur while moving the prepared inputs. The 89 MiB cap is a rounded-up comparison to the separately traced native dynamic-allocation peak, with framework-external allocations still outside the PyTorch cap. These outcomes characterize complete-block residency with the math backend.

The FLASH block control uses a compatibility adapter that checks each of the four masks is floating-point, finite, and identically zero before omitting it from the attention call. Under this mask convention, the official block completes at a 2,048 MiB allocator cap: peak framework allocation is 1,231.55 MiB and sampled process NVML is 1,818 MiB. Table~\ref{tab:block_fidelity} reports its numerical difference from the successful MATH reference alongside the Modified PSI native block. Backend selection therefore changes both memory use and floating-point outputs for this input.

\begin{table}[t]
\caption{Successful block controls on identical frozen inputs. Relative $L_2$ uses the official MATH block as reference. Memory is sampled process NVML. FLASH uses verified zero-mask elision.}
\label{tab:block_fidelity}
\centering\footnotesize\setlength{\tabcolsep}{3pt}
\begin{tabular}{lrrr}
\toprule
Implementation & Peak (MiB) & Video error & Audio error \\
\midrule
Official MATH & 8360 & 0 & 0 \\
Official FLASH adapter & 1818 & 0.002246 & 0.002441 \\
Modified PSI native (80/8) & 360 & 0.003195 & 0.001501 \\
\bottomrule
\end{tabular}
\end{table}

\subsection{Full-pipeline physical-budget control}

Both paths complete the fixed spokesperson case under a 3 GiB physical free pool and produce successfully decoded 121-frame, 704$\times$448 audiovisual files with 48 kHz stereo audio (Table~\ref{tab:pipeline_budget}). Both paths have a sampled workload-process NVML peak of 1,916 MiB, excluding the reservation helper. The native and FLASH pipeline times are 610.37 and 459.59 s. These are single observations of this prompt and seed under the same fixed-reservation procedure.

\begin{table}[t]
\caption{Full-pipeline A6000 control under a fixed 3 GiB physical free pool. Both paths produce successfully decoded audiovisual output. Wall time includes runtime construction, final validation, and eleven 1.5 s measurement pauses; reservation setup is excluded. Peak is sampled workload-process NVML, excluding the reservation helper.}
\label{tab:pipeline_budget}
\centering\footnotesize\setlength{\tabcolsep}{3pt}
\begin{tabular}{lrr}
\toprule
Path & Wall (s) & Peak (MiB) \\
\midrule
Official FLASH & 459.59 & 1916 \\
Modified PSI native & 610.37 & 1916 \\
\bottomrule
\end{tabular}
\end{table}

At 3 GiB, all twelve saved native states and the raw RGB and natural audio outputs exactly reproduce the same unconstrained native case. Against the shared-conditioning MATH reference, native and FLASH final video/audio relative $L_2$ errors are 0.3893/0.05439 and 0.3608/0.06663. Their encoded-video PSNR/SSIM values are 30.53 dB/0.9427 and 30.35 dB/0.9478. Table~\ref{tab:budget_fidelity} in Appendix~\ref{app:budget_fidelity} gives the direct comparison and waveform errors.

\subsection{Matched full-prompt media}

We evaluate three distinct properties. Numerical fidelity measures tensor differences from the matched reference. Media similarity uses encoded-video PSNR and SSIM as image-similarity proxies. Prompt adherence measures whether generated content satisfies the prompt's criteria, with independent ASR for requested speech. Complete audiovisual decoding establishes file validity; similarity and adherence scores characterize the decoded content.

All 27 matched full-prompt outputs pass complete video/audio decoding, with 121 frames at 704$\times$448 and 48 kHz stereo audio. Against the same-case official-transformer endpoint decoded by the common native decoder, Modified PSI has higher encoded-video PSNR in 5/9 cases and higher SSIM in 4/9 cases than baseline PSI. Airliner run \#3 decreases from 23.92 to 19.70 dB PSNR and from 0.8351 to 0.7741 SSIM.

Two fresh \texttt{deepseek-flash} contexts assess all 121 frames against each prompt's five original criteria; all-five-clear requires a score of 2 on every criterion. Context A returns 5/9, 4/9 and 3/9 for official-transformer, baseline PSI and Modified PSI media; context B returns 6/9 for each arm. The contexts agree on 21/27 strict decisions and 119/135 criterion scores. Neither context marks any airliner clip all-five-clear. Both assign airliner run \#3's takeoff criterion 1 for Modified PSI and 2 for the other arms. All nine independent dialogue ASR transcripts exactly match the requested four-word sentence after the fixed normalization. Scores were finalized before implementation identities were revealed.

\subsection{Spark service delivery}

On DGX Spark GB10, requests execute one at a time; the suite covers ten prompt categories in each of three seeded runs. All 30 outputs are verified unchanged from the original service outputs and pass container probing and complete video/audio decoding. Decoded-output throughput uses the three run makespans; mean pipeline time and throughput appear in Table~\ref{tab:service}, with dispersion statistics in Appendix~\ref{app:service}.

\begin{table}[t]
\caption{Single-slot DGX Spark GB10 service results. Pipeline duration covers each job's generation; throughput uses the complete orchestration spans of the three runs.}
\label{tab:service}
\centering\footnotesize
\begin{tabular}{lr}
\toprule
Metric & Spark GB10 \\
\midrule
Runs $\times$ requests & $3\times10$ \\
Active execution setting & Single slot \\
Decoded outputs / requests & 30/30 \\
Mean pipeline time (s) & 524.143 \\
Decoded outputs/GPU-hour & 6.862 \\
All-five-clear scores (A/B) & 8/30; 14/30 \\
\bottomrule
\end{tabular}
\end{table}

The Spark configuration uses BF16 without quantization, a 1,024-token dynamic context, and an 80 MiB arena. Its unified-memory accounting is reported separately from the RTX A6000 process NVML measurements.

\subsection{Spark prompt-adherence evaluation}

We assess each of the 30 successfully decoded Spark outputs against the five criteria in its original prompt. Each API request contains all 121 original frames at 704$\times$448, their timestamps, and independent Whisper small.en transcripts for speech prompts. Hardware and prior scores are hidden. Two fresh \texttt{deepseek-flash} contexts use the same frozen rules with different instruction and criterion order. Each criterion receives 0 (absent or contradicted), 1 (partial or ambiguous), 2 (clearly supported), or an unassessable label. An all-five-clear result requires five scores of 2.

The two contexts return 8/30 and 14/30 all-five-clear results. Criterion scores agree on 134/150 items, and the all-five-clear decisions agree on 24/30 clips. These repeated model-assisted evaluations disagree on borderline prompt-adherence judgments.

Independent ASR exactly matches the requested dialogue after case and punctuation normalization for all 6/6 Spark speech clips. ASR measures recovered words; speaker attribution and lip synchronization require separate audiovisual assessment. Appendix~\ref{app:semantic} specifies the evaluation protocol.

\section{Related work and positioning}

The residency unit determines which memory constraint offloading can relieve. In training, vDNN manages activation lifetimes across CPU and GPU, while SwapAdvisor jointly selects operator order, allocation, and tensor swaps \cite{vdnn,swapadvisor}. ZeRO-Infinity explicitly decomposes oversized operators into parameter tiles fetched and released sequentially \cite{zeroinfinity}. For autoregressive inference, FlexGen optimizes tensor placement and access over GPU, CPU, and disk \cite{flexgen}. These precedents distinguish total model storage from the parameters and live state needed for the current computation.

Limiting that live state also requires scheduling intermediates and data movement within and across operators. Welder's tile graphs coordinate intra- and inter-operator reuse across a memory hierarchy that can include the host \cite{welder}. FlashAttention reduces HBM--SRAM traffic through tiled exact attention \cite{flashattention}, and Blockwise Parallel Transformers carries sequence-block attention into feed-forward computation \cite{bpt}. Operator tiling, intermediate reuse, and online attention are therefore established ingredients of memory-efficient execution.

Diffusion systems combine these choices under latency and memory budgets. Streamlined Inference uses feature slicing and operator grouping for activation memory, with a separate step-reuse mechanism \cite{streamlined}. Xema couples selective block-weight offload with localized activation chunking and fusion and static lifetime-based memory placement \cite{xema}. ChunkFlow splits layer prefetch to yield to distributed collectives and permits partial parameter residency \cite{chunkflow}; its transfer chunks refine communication scheduling while computation proceeds layerwise. Diffusers exposes block- and leaf-level offload with stream prefetch \cite{diffusersoffload}, and WeeLLM documents unquantized layer streaming and lists LTX-2.5 support \cite{weellm}. Lightricks' reference pipeline also documents FP8 casting with CPU or disk offload for constrained GPU memory \cite{ltx25runtime}.

PSI organizes operator partitioning and host-backed state for the BF16 LTX-2.5 22B audiovisual transformer \cite{ltx25model,ltx2} using fixed-size weight pages, separately indexed host state, and explicit producer/consumer dependencies. Its contribution is this construction and its validation: a managed-arena bound with stated state and workspace assumptions; separate arena, process, and complete-pipeline measurements; and numerical and media checks that distinguish real-arithmetic identities, transport-only byte preservation, BF16 execution differences, and successfully decoded audiovisual outputs. The fixed-budget controls include both PSI and a complete-block FLASH path.

\section{Conclusion}

PSI executes a 22B BF16 joint audio--video transformer by scheduling weight slices through a fixed arena and retaining intermediate state in indexed host buffers. Its partitioned operators preserve the dense real-arithmetic construction under the stated assumptions. Event-ordered transport preserves operand bytes; the fixed one-token control reproduces twelve saved states bitwise.

Across eighteen full-prompt native trajectories on RTX A6000, both denoising stages use an 80 MiB managed arena and reach a sampled 360 MiB transformer-process peak. Complete native generation and the complete-block FLASH control both succeed under a fixed 3 GiB available-memory cap. The matched three-path study produces 27/27 successfully decoded audiovisual outputs, and the separate Spark service suite produces 30/30. These results distinguish transformer residency, complete-pipeline memory, and decoded-output validity.

Finite-precision evaluation accounts for the same-input projection discrepancies. Targeted bias and attention changes provide local numerical improvements and lower both one-token endpoint errors; across nine full-prompt cases, mean error increases in both modalities. With arithmetic fixed, FFN reuse lowers mean block time by 4.0\% in the five-repeat scheduling study. The construction thus supports small managed residency with explicit scheduling and numerical tradeoffs. The matched media results cover three prompt categories with common conditioning and decoders; broader perceptual evaluation and localization of the remaining trajectory differences are directions for further study.

\appendices
\raggedbottom
\section{Supporting measurements and diagnostics}
\label{app:provenance}

Baseline PSI includes the Q/K normalization and rotary-position materialization changes evaluated at denoising state 5 in the first transformer block. Modified PSI adds the bias and attention changes described in the numerical comparison. The Spark service suite is separate from the RTX A6000 matched full-prompt and fixed-budget experiments.

The 17 native source files retained on Spark are byte-identical to the corresponding baseline copies. Spark records 330 successful native transformer invocations, eleven per job, each completing 48 blocks. The available implementation combines FFN reuse and asynchronous journal transport; it uses the compute stream for weight copies, so it does not include the separate weight-prefetch optimization.

\subsection{Service records and timing distributions}
\label{app:service}

Spark executes one request at a time. All 30 original MP4 files pass container and stream probing and full video/audio decoding to end of file. Validation uses FFmpeg 6.1.1 configured to stop on decoding errors; no errors occur. Every file contains 121 video frames at 704$\times$448 and 24 fps, with a 5.041667 s video/container duration and 48 kHz stereo AAC audio. The media were validated on 30 October 2026.

The monotonic pipeline timer starts after setup and ends after decoding and multiplexing; it excludes time waiting in the service queue. These pipeline times have a sample standard deviation of 4.419 s and a linearly interpolated P95 of 529.961 s. Each job includes eight Stage-1 and three Stage-2 transformer invocations. Spark's throughput uses the 30 decoded outputs and the three measured orchestration spans: 5261.613492, 5231.609223, and 5246.526462 s. These include polling and orchestration overhead beyond the worker intervals.

\subsection{Semantic evaluation protocol}
\label{app:semantic}

The Spark evaluation covers all 3,630 frames from the 30 decoded videos and uses 60 separate API calls made on 30 October 2026. The API identifies the model as \texttt{deepseek-flash}; service documentation identifies that deployment as DeepSeek-V4.1-Flash-1017. No previous judgment is supplied to a later call. Evaluation records retain the input prompts, raw responses, and timestamped evidence for each criterion. Both contexts use the same model.

Whisper small.en runs independently in CPU INT8 mode with no requested-text prompt. Across the six Spark speech clips, the transcripts have zero word edits among 36 requested words.

\subsection{Fixed-budget state and media comparisons}
\label{app:budget_fidelity}

Table~\ref{tab:budget_fidelity} compares the completed 3 GiB paths with the same-case unconstrained MATH transformer and with each other. All use common native conditioning, decoding, and decoder noise. Relative $L_2$ uses the named reference's norm. RGB metrics cover all 114,487,296 decoded BF16 values before clipping or encoding; audio metrics cover 240,480 natural samples per stereo channel without alignment, normalization, or mux padding. The direct FLASH-to-native encoded comparison gives 29.92 dB PSNR and 0.9471 SSIM over all 121 frames. The constrained native rerun matches all twelve same-native unconstrained states and raw RGB/audio samples exactly; its encoded frames have zero error. Per-state, per-frame, and per-channel records accompany the retained inputs and outputs.

\begin{table}[t]
\caption{Relative $L_2$ in the completed fixed-budget case. MATH denotes the same-case unconstrained MATH transformer with the common native decoder; native denotes the 3 GiB Modified PSI path.}
\label{tab:budget_fidelity}
\centering\footnotesize\setlength{\tabcolsep}{3pt}
\begin{tabular}{lrrrr}
\toprule
Candidate / reference & Video & Audio & RGB & Waveform \\
\midrule
Native / MATH & 0.3893 & 0.05439 & 0.1658 & 0.5874 \\
FLASH / MATH & 0.3608 & 0.06663 & 0.1684 & 0.8468 \\
FLASH / native & 0.3637 & 0.07136 & 0.1796 & 0.8249 \\
\bottomrule
\end{tabular}
\end{table}

\begin{samepage}
\section*{Acknowledgment}
Hanyue Shen and Xinling Liao contributed equally to all aspects of the work. The authors declare no conflicts of interest. OpenAI Codex assisted with implementation diagnostics, evidence analysis, and manuscript preparation. DeepSeek supplied the model-assisted semantic judgments described above. The authors are responsible for the technical content and final manuscript.

\end{samepage}

\IEEEtriggeratref{11}
\bibliographystyle{IEEEtran}
\bibliography{references}
\end{document}